\documentclass{aa}
\usepackage{graphics}
\def\a{{$\alpha$}}

\def\gsnr{{G~82.2+5.3}}
\newcommand{\h}{$^{\rm h}$}
\newcommand{\m}{$^{\rm m}$}
\newcommand{\s}{$^{\rm s}$}
\newcommand{\dd}{$\delta$}
\newcommand{\ha}{\rm H$\alpha$}
\newcommand{\hbeta}{\rm H$\beta$}
\newcommand{\HII}{\ion{H}{ii}}
\newcommand{\hnii}{{\rm H}$\alpha+[$\ion{N}{ii}$]$}
\newcommand{\nii}{$[$\ion{N}{ii}$]$}
\newcommand{\sii}{$[$\ion{S}{ii}$]$}
\newcommand{\oi}{$[$\ion{O}{i}$]$}
\newcommand{\he}{$\ion{He}{i}$}
\newcommand{\oii}{$[$\ion{O}{ii}$]$}
\newcommand{\oiii}{$[$\ion{O}{iii}$]$}
\newcommand{\snr}{\rm supernova remnant}

\newcommand{\et}{et al.}
\newcommand{\flux}{$10^{-17}$ erg s$^{-1}$ cm$^{-2}$ arcsec$^{-2}$}
\newcommand{\dens}{\rm cm$^{-3}$}
\newcommand{\sdens}{\rm cm$^{-2}$}
\newcommand{\vel}{\rm km s$^{-1}$}
\newcommand{\dof}{{\it dof}}

\begin{document}

%   \thesaurus{08     % A&A Section 8: Diffuse matter in space
%              (09.07.01;  % ISM : general,
%		09.19.2;
%		09.09.1)} % Superona remnants
%
\title{Multi-wavelength study of the \gsnr\ supernova remnant}
\author{F. Mavromatakis\inst{1}
\and B. Aschenbach\inst{2}
\and  P. Boumis\inst{1,}\thanks{Present address:Institute of Astronomy \& Astrophysics, National Observatory of Athens, I. Metaxa \& V. Pavlou, P. Penteli, 15236 Athens, Greece}
\and J. Papamastorakis\inst{1,3}
}
\offprints{F. Mavromatakis,\email{fotis@physics.uoc.gr}}
\authorrunning{F. Mavromatakis et al.}
\titlerunning{Observations of \gsnr}
\institute{
University of Crete, Physics Department, P.O. Box 2208, 710 03 Heraklion, Crete, Greece 
\and Max-Planck-Institut für extraterrestrische Physik, Postfach 1312, 
D-85741 Garching, Germany
\and Foundation for Research and Technology-Hellas, P.O. Box 1527, 711 10 Heraklion, 
Crete, Greece
}
%\date{Received ...... / Accepted .......}
\date{Received 8 August 2003/Accepted 6 November 2003}

\abstract{
We present the first CCD flux--calibrated images of the supernova remnant 
\gsnr\  in major optical emission lines. The medium ionization line of 
\oiii 5007 \AA\ provides the first direct evidence of optical emission 
originating from \gsnr. 
Filamentary emission is detected in the west and east areas of the
remnant, roughly defining an ellipsoidal shell. The \oiii\ emission is
rather well correlated with the radio emission suggesting their association, 
while typical fluxes are found in the range of 20--30$\times$\flux. 
Deep long--slit spectra taken at specific positions of the remnant verify 
that the detected filamentary emission originates from shock heated 
gas, while the diffuse \oiii\ emission in the south results from 
photoionization processes. The spectra further suggest shock velocities 
around 100 \vel\ and low electron densities. The X--ray surface brightness is 
quite patchy, 
missing obvious limb brightening and is dominated by a bright   
bar--like emission region which is off-set from the geometric center 
by $\sim$9\arcmin. 
The X-ray emission is thermal and requires two temperatures of 0.2 keV and 
0.63 keV. 
The bright bar region shows overabundant Mg, Si and Fe,  
which might indicate still radiating ejecta matter. 
The azimuthally averaged radial surface profile is consistent with the matter 
density changing 
with distance r from the center $\propto$e$^{\rm -r/r_0}$ with a characteristic 
angular length of 36\arcmin, or, alternatively, with an r$\sp{-1/2}$ density profile. 
The matter inside the remnant is   
quite likely structured like a porous cloudy medium. The average matter 
density is $\sim$0.04$\times{\rm d\sb{1.6}\sp{-0.5}}$ with 
d$\sb{1.6}$ the distance in units of 1.6 kpc. Because of the low density and 
the long cooling times involved the remnant is more likely to be in the 
adiabatic phase, which is consistent with the densities derived for the X-ray 
plasma and the optical line emission, but it is not excluded that is has 
reached the radiating phase. This, however, would imply a lower density, greater 
age and much larger distance, at the edge of the upper limits obtained from 
N$\sb{\rm H}$ and E(B-V).
}
\maketitle
\section{Introduction}
The supernova remnant \object{G 82.2+5.3} (\object{W 63}) is found in the 
complex region of the Cygnus constellation. 
Radio observations at several frequency bands established the nature 
of the object through its non--thermal radio spectrum  
(Velusamy \& Kundu \cite{vel74}, Angerhofer \et\ \cite{ang77}, 
Wendker \cite{wen71}). The remnant  
displays a non--circular, elliptical shape of $\sim$70\arcmin$\times$100\arcmin\ 
in the radio map at 11 cm. The distance to the remnant is not well determined 
and distances in the range of 1.3--1.9 kpc have been 
reported (Rosado \& Gonzalez \cite{ros81} and references therein).
Interference--filter photographs from Parker \et\ (\cite{par79}) revealed  
bright extended  structures in \ha\ and \sii.
There is no obvious correlation between the low ionization 
images and the radio maps of \gsnr. Optical spectra taken from Sabbadin 
(\cite{sab76}) show the characteristic signature of emission from shock 
heated gas (\sii/\ha $\sim$0.5). 
Rosado \& Gonzalez (\cite{ros81}) estimated expansion velocities for some 
of the optical filaments of $\sim$35 \vel\ and $\sim$70 \vel, depending upon 
the size assumed for the remnant. 
\par
The X-ray and radio morphologies of the remnant are quite different. 
Rho \& Petre (\cite{rho98}) proposed that it belongs to the class of 
mixed--morphology remnants. Indeed, the X-ray emission is neither 
shell-like nor plerion-like but the 
vast majority of the X--rays originate from areas around but not peaked on 
the remnant center, and there is no obvious limb brightening. 
The entire X-ray region is roughly 
bound by the radio shell. Rho \& Petre (\cite{rho98}) analysing ROSAT data 
describe the entire spectrum by a thermal, single temperature spectrum with 
kT $\sim$0.2 keV subjected to interstellar absorption with a column density of 
$\sim$4 $\times$ 10$^{21}$ \sdens\, although with large uncertainties.
\par
Flux calibrated images in major optical emission lines were obtained and 
deep long--slit spectroscopy of selected areas of interest was performed for 
a detailed study of the optical emission in the area of \gsnr.
In addition, we analyzed ROSAT All--Sky survey and re--analyzed ROSAT pointed 
data as well as ASCA GIS and SIS data from one pointing.  
Information about the observations and the data 
reduction is given in Sect. 2. In Sect. 3 and 4 we present the results of 
our imaging and spectral observations. Finally, in Sect. 5 we discuss 
the properties of the remnant and its environment.
\section{Observations}
\subsection{Optical images}
The current wide field observations were performed with the 0.3 m 
Schmidt--Cassegrain telescope at Skinakas Observatory, Crete, Greece. 
The complex field of \gsnr\ was observed on July 25 and 26, 
2001. The 1024 $\times$ 1024 SITe CCD used during these imaging 
observations resulted in a 89\arcmin\ $\times$ 89\arcmin\ field of view and an 
image scale of 5\arcsec\ per pixel. 
A journal of the observations together with information about the filters 
are given in Table~\ref{obs}. 
The images presented in this work are the average of the available frames 
from the individual filters. 
The astrometric solutions for all data frames utilized 
the HST Guide star catalogue (Lasker \et\ \cite{las99}) and all coordinates 
quoted in this work refer to epoch 2000. All frames were projected to a 
common origin on the sky.
\par
Standard IRAF and MIDAS routines were employed for the reduction of the data. 
All frames were bias subtracted and flat-field corrected using a series of
well exposed twilight flat--fields. The spectrophotometric standard stars 
HR5501, HR7596, HR7950 and HR9087 were observed in order to obtain 
absolute flux calibration (Hamuy \et\ \cite{ham92}, \cite{ham94}).
\begin{table}
      \caption[]{Journal of the Observations}
         \label{obs}
\begin{flushleft}
\begin{tabular}{lllll}
            \noalign{\smallskip}
\hline
   &   $\lambda_{\rm C}$  & $\Delta\lambda$  &   & Exp. times$^{\rm a}$ \cr 
Filter &    (\AA)  &   (\AA)  & Date (UT) &   (No of frames$^{\rm b}$) \cr 
\hline
\hnii   &    6560               & 75       & 25--26/07/2001& 7200 (3)\cr
\sii    &    6708               & 20       & 25/07/2001    & 7200 (3)  \cr
\oiii   &    5005               & 28       & 26/07/2001    & 9600 (4) \cr
\oii    &    3727               & 25       &  26/07/2001   & 7200 (3)  \cr
Cont red  & 6096               & 134      & 25--26/07/2001 & 720 (8)\cr
Cont blue & 5470               & 230      & 26/07/2001    & 360 (2) \cr
\hline
\end{tabular}
\end{flushleft}
${\rm ^a}$ Total exposure times in sec \\\
${\rm ^b}$ Number of individual frames \\\
\end{table}
  \begin{table}
      \caption[]{Spectral log}
         \label{spectra}
\begin{flushleft}
\begin{tabular}{lllll}
            \noalign{\smallskip}
\hline
	Slit centers &  & Exp. times$^{\rm a}$ & \cr
\hline
	 $\alpha$ & $\delta$ & (No of spectra$^{\rm b}$)  &  \cr
\hline 
 20\h21\m39\s.3 & 45\degr59\arcmin30\arcsec.1 	& 16500 (5) \cr
\hline
 20\h20\m04\s.0 & 44\degr54\arcmin48\arcsec.5 	& 7200 (2)\cr
\hline
 20\h16\m15\s.6 & 45\degr51\arcmin58\arcsec.9 	& 7200 (2) \cr
 \hline
 20\h16\m15\s.4 & 45\degr58\arcmin01\arcsec.4 	& 7200 (2)\cr
 \hline
\end{tabular}
\end{flushleft}
${\rm ^a}$ Number of spectra obtained \\\
${\rm ^b}$ Total exposure times in sec\\\
   \end{table}
  \begin {figure*}
   \resizebox{\hsize}{!}{\includegraphics{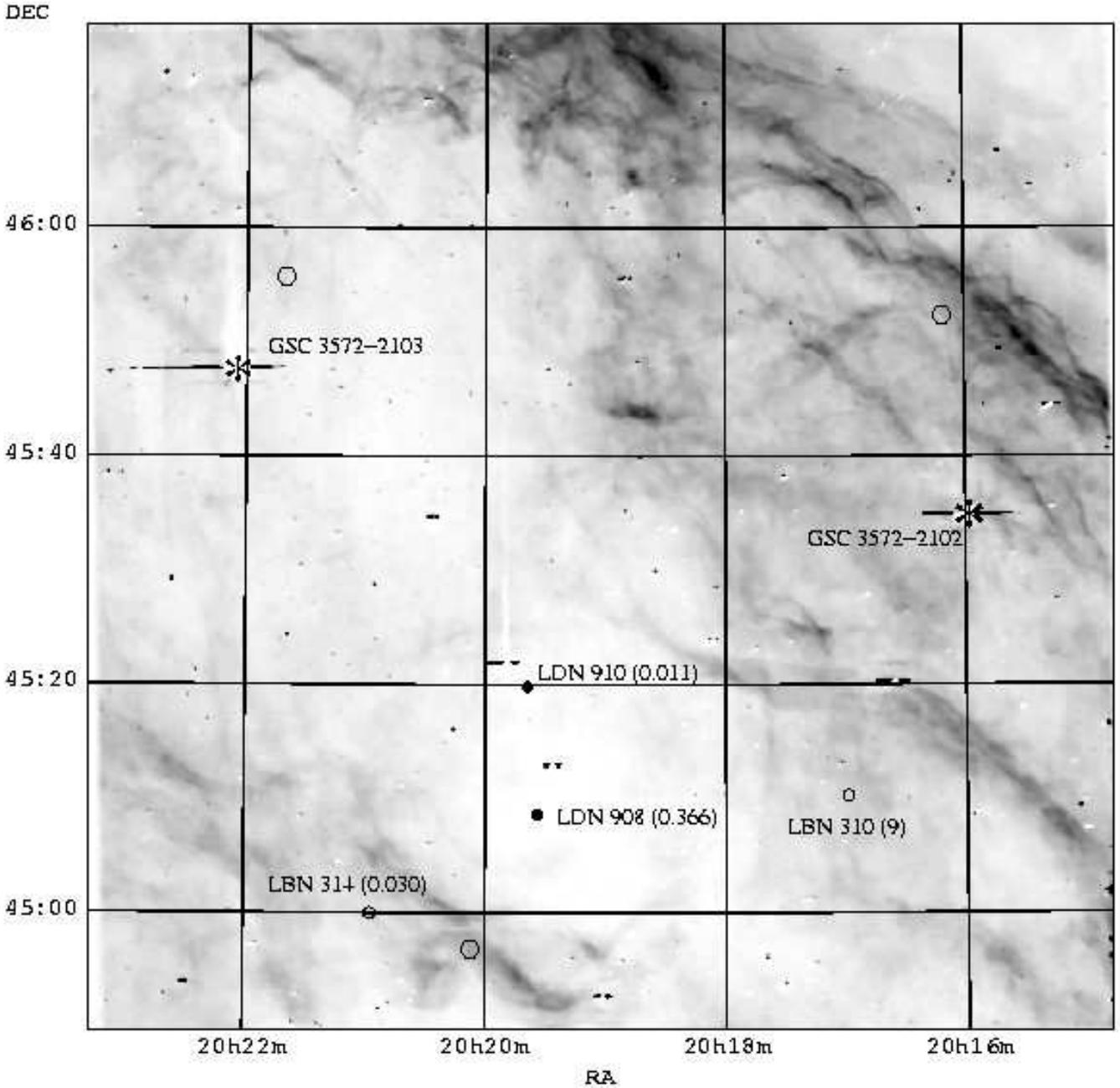}}
    \caption{ The field of \gsnr\ in the \hnii\ filter. 
     The image has been smoothed to suppress the residuals 
     from the imperfect continuum subtraction. The open and the filled
     circles designate the positions of bright and dark nebulae, 
     respectively, while the numbers in parentheses represent the 
     surface of each nebula in square degrees as given by 
     Lynds (\cite{lyn62}, \cite{lyn65}). The open polygons show the
     positions where emission line ratios are measured (see also \S 3.3).
     The shadings run linearly from 0 to 400 $\times$ \flux. 
     The line segments seen near over--exposed stars in this 
     figure and the next figures are due to the blooming effect.
      } 
     \label{fig01}
  \end{figure*}
%--------------------------------------------------------
  \begin {figure*}
   \resizebox{\hsize}{!}{\includegraphics{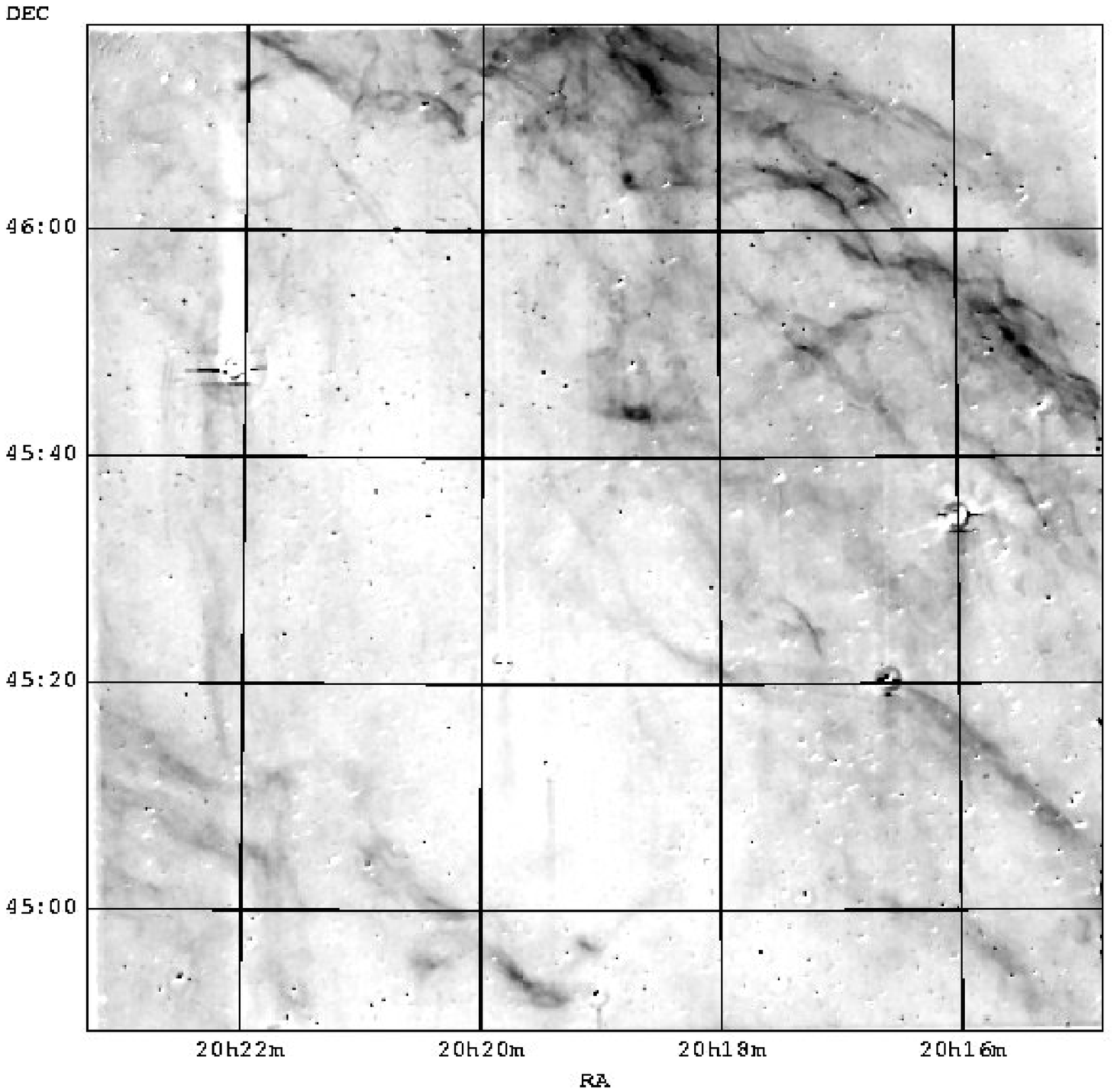}}
    \caption{ The \sii\ image of the area around \gsnr. The morphology 
    in these emission lines is similar to that of the \hnii\ lines. 
     The image has been smoothed to suppress the residuals 
     from the imperfect continuum subtraction, while the
     shadings run linearly from 0 to 60 $\times$ \flux. 
     } 
     \label{fig02}
  \end{figure*}
\subsection{Optical spectra}
The 1.3 m Ritchey--Cretien telescope at Skinakas Observatory was used 
to obtain long--slit spectra on June 21, 22, and July 21, 22 2001. 
The data were taken with a 1300 line mm$^{-1}$ grating 
and a 800 $\times$ 2000 SITe CCD covering the range of 4750 \AA\ -- 6815 \AA.
The slit had a width of 7\farcs7 and a length of 7\arcmin.9 and, in all cases, 
was oriented in the south-north direction. The coordinates of the slit centers, the 
number of available spectra from each location and the total exposure time of 
each spectrum are given in Table~\ref{spectra}. 
The spectrophotometric standard stars HR5501, HR7596, HR9087, HR718 
and HR7950 were observed for the absolute calibration of the spectra of \gsnr.  
%
%-----------------------------------------------------------------
  \begin{table*}
        \caption[]{Relative line fluxes}
         \label{sfluxes}
         \begin{flushleft}
         \begin{tabular}{lllllll}
     \hline
 \noalign{\smallskip}
                & I (west)   & II (east)  & III (south) \cr
\hline
Line (\AA) & F$^{\rm 1,2}$ & F$^{\rm 1,2}$ & F$^{\rm 1,2}$   \cr
\hline
4861 \hbeta\    & 25 (19) &  15 (10)  &  11.4 (66)   \cr
\hline
4959 \oiii$_1$ & 34 (20)  & 159 (104) &  1.5  (11)  \cr
\hline
5007 \oiii$_2$ & 105 (57) & 511 (293) & 5.0 (36)   \cr
\hline  
5872 \he      &  --      &  --       & 3.5 (37)   \cr
\hline
6300 \oi$_1$  &   --     &  75$^3$ (32)  &   --    \cr
\hline
6364 \oi$_2$  &  --      & 22$^3$ (13)   &   --    \cr
\hline
6548 \nii$_1$ & 15 (18)  &  44 (44)  &  9.7 (95)   \cr
\hline
6563 \ha\     & 100 (65) & 100 (73)  & 100 (520)   \cr
\hline
6584 \nii$_2$ & 63 (55)  & 141 (115) & 32.2 (260)  \cr
\hline
6678 \he     &   --      &  --      & 1.3 (17)     \cr
\hline
6716 \sii$_1$ & 35 (39)  & 78 (73)   & 10.8 (113)  \cr
\hline
6731 \sii$_2$ & 25 (29)  & 56 (53)   & 7.6 (82)    \cr
\hline
\hline
Absolute \ha\ flux$^{\rm 4}$ & 15.0 &   4.3  & 72  \cr
\hline
\ha /\hbeta\     & 4.0 (18)   & 6.7 (10) &   8.8 (65)     \cr
\hline
\oiii /\hbeta    & 5.6 (18)  & 45 (10)   &   0.57 (30)    \cr 
\hline
\sii/\ha\        & 0.6 (39)  & 1.34 (56) &   0.18 (44)    \cr
\hline 
I(6716)/I(6731)& 1.40 (23) & 1.39 (43) &  1.42 (66)     \cr
\hline 
\hline
c$^{\rm 5}$    & 0.38 [0.07] & 1.1 [0.1]&   1.41 [0.01]  \cr
\hline
E(B-V)         & 0.25[0.05]  & 0.73[0.07] & 0.94[0.01]    \cr
\hline
n$_{\rm e}$    & $<$ 80 \dens & $<$ 30 \dens & $<$ 30 \dens & \cr
\hline
V$_{\rm s}$     & $<$ 100 \vel   & $>$ 100 \vel & -- \cr
\hline
\end{tabular}
\end{flushleft}
 ${\rm ^1}$ Uncorrected for interstellar extinction 

${\rm ^2}$ Listed fluxes are a signal to noise weighted
average of the individual fluxes

$^{\rm 3}$ Present in only one of the spectra

$^{\rm 4}$ In units of \flux\ 

$^{\rm 5}$ The logarithmic extinction is determined as 
c = 1/0.331 $\cdot$ log((\ha/\hbeta)$_{\rm obs}$/3)

${\rm }$ Numbers in parentheses represent the signal to noise ratio 
of the quoted fluxes \\
${\rm }$ The 1$\sigma$ error is given in square brackets for the logarithmic 
extinction\\
${\rm }$ All fluxes normalized to F(\ha)=100
\end{table*}
%-------------------------------------------------------
%
\begin{table*}
   \caption[]{X--ray spectral fits}
    \label{bestfit}
\begin{flushleft}
\begin{tabular}{llllll}
            \noalign{\smallskip}
\hline
Instrument&N$\sb{\rm H}$          &kT                        &EM                      &$\chi\sp 2\sb{\rm{red}}$&\it{dof}\cr
          &10$\sp{21}$cm$\sp{-2}$ &keV                       &10$\sp{-3}$cm$\sp{-5}$  &                        &         \cr
\hline
GIS/SIS  
&4.7$\sp{+0.5}\sb{-0.4}$&0.63$\sp{+0.05}\sb{-0.04}$&1.50$\sp{+0.20}\sb{-0.20}$&0.99                    &198\cr
PSPC$\sp
1$&0.8$\sp{+6.4}\sb{-0.6}$&0.67$\sp{+0.10}\sb{-0.47}$&0.57$\sp{+0.30}\sb{-0.30}$&0.74                    &22\cr
PSPC$\sp 2$&4.0$\sp{+4.0}\sb{-1.8}$&0.21$\sp{+0.06}\sb{-0.04}$&4.70$\sp{+4.7}\sb{-2.1}$&0.79                    &48\cr
\hline
\end{tabular}
\end{flushleft}
${\sp 1}$: one temperature vmekal model\\\
${\sp 2}$: two temperature vmekal model; values of low temperature  \\\ 
component listed, values of high temperature component \\\ 
restricted to GIS/SIS fit range.\\\
\end{table*}
\subsection{The ROSAT and ASCA observations}
The Cygnus region was scanned during the ROSAT All-sky survey in early December 
1992. X-rays from G 82.2+5.3, which is located at the northern tip of the 
so-called Cygnus Superbubble shell (Uyaniker \et\ \cite{uya01}) were clearly 
detected and an image of the remnant and the environment could be constructed. 
The net observing time was $\sim$1200 s. 
The remnant under study was also observed several times with the ROSAT PSPC in pointing mode, 
the longest observation of which was for $\sim$8 ks on 26 October 1993 
aiming at the center of the remnant (sequence $\#$500286p). 
Additional ROSAT observations were targeted at the west of the remnant ($\#$500219p/500219p1), 
the north-east ($\#$500217p) and the south-east ($\#$500218p). Apart from the pointing 
towards the centre 
only the south-east pointing provided sufficient counts to create a spectrum over the central 
40 arcmin of the PSPC. We also found one observation in the ASCA archive and both the GIS and SIS data 
could be used for a spectral analysis. 
\section{The optical emission}
\subsection{The \hnii\ and \sii\ line images}
The images in \hnii\ (Fig. \ref{fig01}) and \sii\ (Fig. \ref{fig02}) of the remnant look
complex due to the presence of several bright and dark nebulae. 
The dark nebulae LDN 908 and 910 (Lynds \cite{lyn62}) are found in the south 
areas of our field, while  LBN 081.88+04.79 (LBN 314; Lynds \cite{lyn65}) is 
located in the south--east and LBN 082.79+05.83 (LBN 325), LBN 083.03+05.78 
(LBN326) located to the north may contribute to the emission in our field. 
Lynds (\cite{lyn65}) also reports LBN 081.62+05.47 (LBN 310) as a very 
extended \HII\ region which seems to overlap our field. 
The complexity of the environment of the remnant  
does not allow the direct identification of any optical emission matching the 
morphology of the known radio emission. However, since our images are flux 
calibrated they can be used to approximately determine the 
emission characteristics. 
The ratio of the low ionization images of \sii\ and \hnii\ shows that 
the bulk of the emission in the west results from photoionization, 
while the weak emission seen to the south of the bright star 
\object{GSC 3572-2102} probably originates from shock heated gas. 
In this area, roughly defined along \a\ $\simeq$ 20\h22\m30\s\ and between 
\dd\ $\simeq$ 45\degr47\arcmin\ and 45\degr12\arcmin, we estimate a \sii/\ha\ ratio of
$\sim$0.7--0.8, while LBN 082.79+05.83 and the bright structure in the 
north--west are characterized by ratios very close to $\sim$0.4.   
\subsection{The \oii\ and \oiii\ images}
The \oii\ image of the area around \gsnr\ is not shown here since  its 
morphology is generally similar to that of the \hnii\ and \sii\ low
ionization images.
Emission from the \HII\ region in the \oii 3727 \AA\ line is somewhat suppressed
but still the shape of the remnant is not clearly outlined.
No immediate correlation of the optical emission with the known radio 
emission can be seen. 
In strong contrast to the low ionization images, the field of \gsnr\ 
appears markedly different in the medium ionization line of \oiii 5007 \AA\ 
(Fig. \ref{fig04}). The overall emission is significantly attenuated and the 
image appears to be free of complex structures. 
Optical radiation possibly from LBN 082.79+05.83 and/or 083.03+05.78 
is seen in the
north edge of the field of view. The \oiii\ image reveals filamentary emission 
in the west and east areas as well as diffuse emission. 
Bright filaments are detected along \a\ $\simeq$ 20\h16\m\ and 20\h22\m\ and 
around \dd\ $\simeq$ 45\degr40\arcmin. These two filamentary structures define 
rather well the opposite sides of an ellipsoidal shell which is in agreement 
with the morphology of the radio emission. 
In fact the correlation of the \oiii\ emission in the east with the 
radio emission observed at 326 MHz and 4850 MHz (Fig. \ref{fig05}) suggests its 
association to \gsnr. Fainter structures are also detected in the south and 
inner regions of the field. Faint emission from a region shaped as an arc is 
detected to the east of GSC~3572--2102, roughly along \a\ $\simeq$ 20\h22\m40\s, 
from \dd\ $\simeq$ 45\degr30\arcmin\ to 46\degr30\arcmin. 
This faint filamentary structure is not continuous but gaps in intensity are 
detected along its extent. This may indicate the presence of inhomogeneities 
in the interstellar ``clouds'' resulting in strong variations of the shock 
velocity. It is known that the shock velocity greatly influences the 
\oiii\ flux (Cox \& Raymond \cite{cox85}). The typical projected angular 
width of the \oiii\ filaments is $\sim$30\arcsec, equivalent to 0.2 pc at 
an assumed average distance of 1.6 kpc (see \S 1). 
\subsection{The long--slit spectra from \gsnr}
\begin {figure*}
   \resizebox{\hsize}{!}{\includegraphics{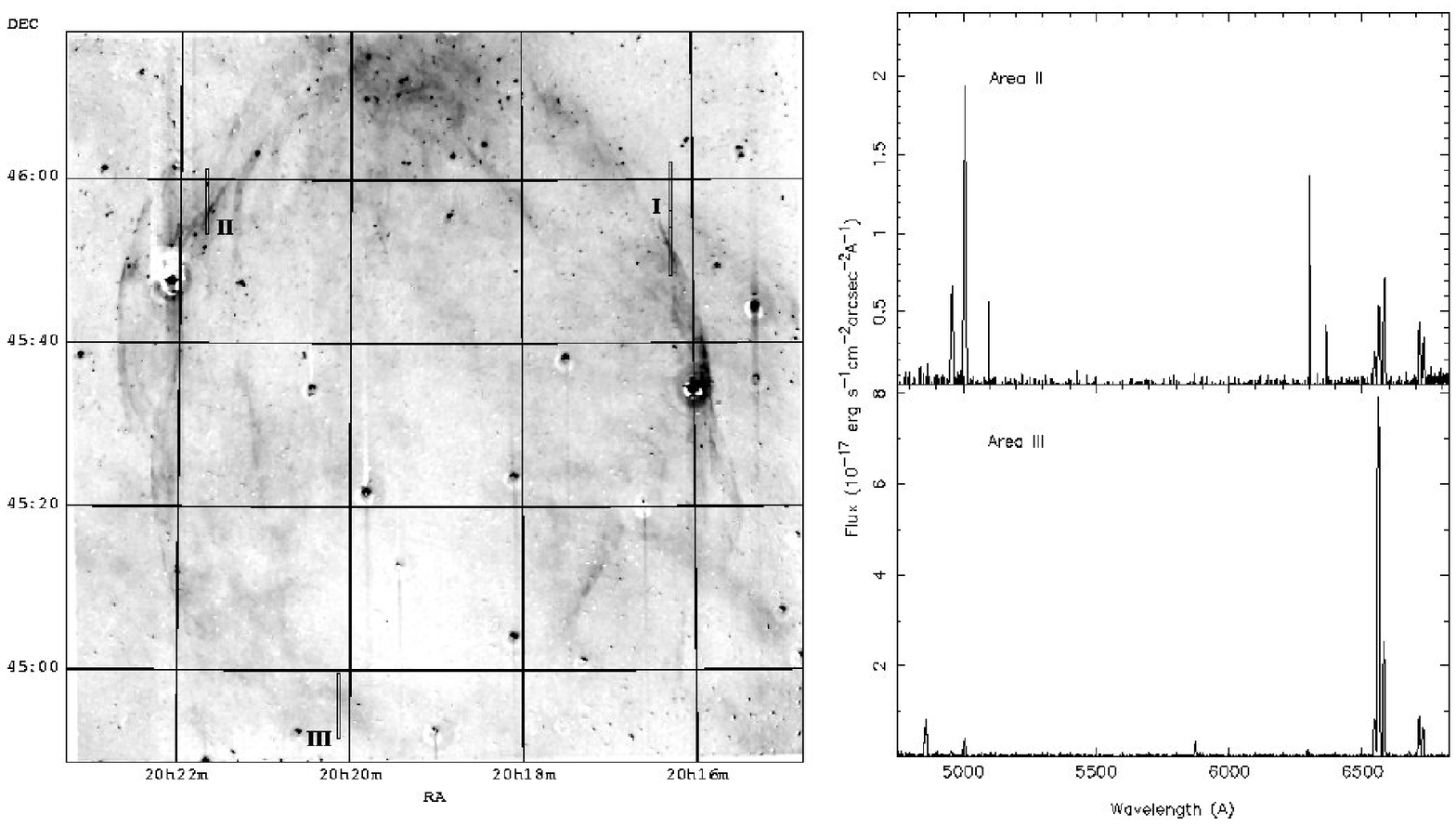}}
    \caption{ The medium ionization line of \oiii 5007 \AA\ provides the 
     sharpest view of \gsnr\ (left).  
     The image has been smoothed to suppress the residuals 
     from the imperfect continuum subtraction and the shadings run 
     linearly from 0 to 30 $\times$ \flux. The long rectangles indicate 
     the projected positions of the slits on the sky and the numbers designate
     the areas discussed in the text. The right figure shows single 
     long--slit spectra from areas II and III. 
     } 
     \label{fig04}
  \end{figure*}

\begin{figure*}
   \resizebox{\hsize}{!}{\includegraphics{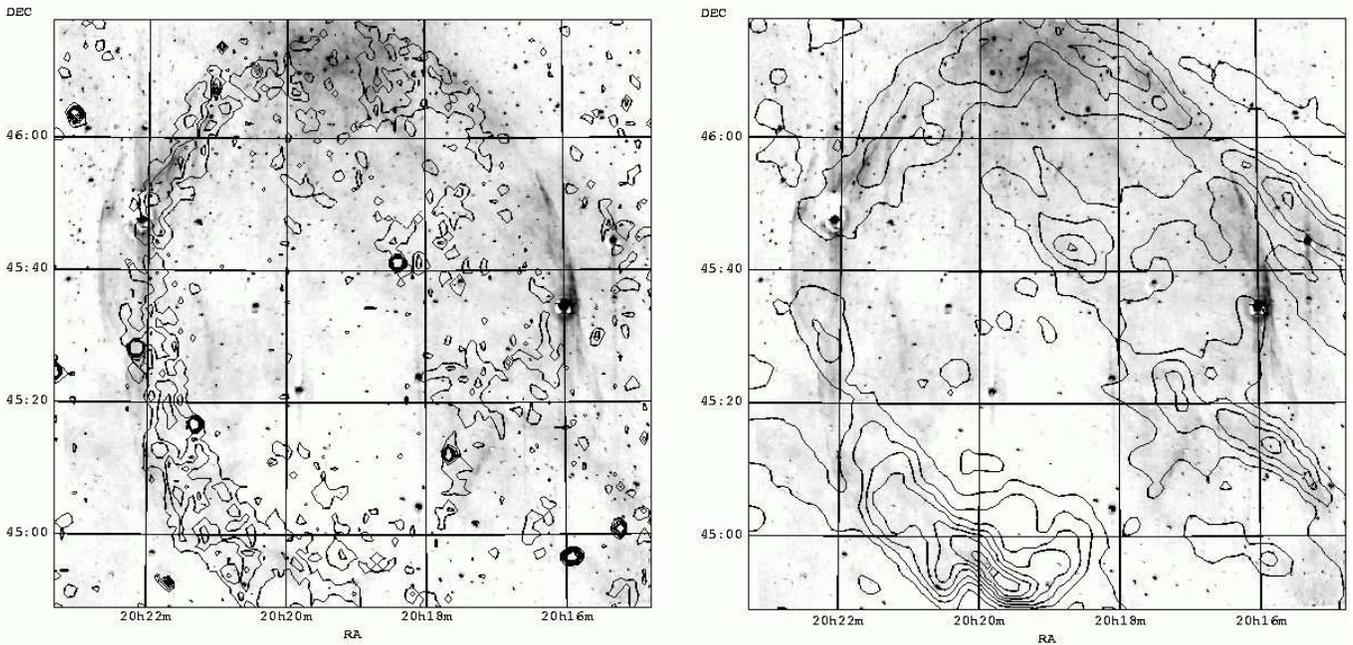}}
    \caption{The correlation between the \oiii\ emission and the radio 
     emission at 326 MHz (left) and 4850 MHz (right) is shown in this figure. 
     The 326 MHz radio contours (Rengelink \et\ \cite{ren97}) scale linearly 
      from 0.01 Jy/beam to 0.09 Jy/beam, while the 4850 MHz contours scale 
      from 5 10$^{-4}$ Jy/beam to 0.2 Jy/beam (Condon \et\ \cite{con94}). 
      The optical and radio data are rather well correlated in the east 
      areas of the remnant. The radio emission in the west seems to be 
      weeker and mainly related to the \HII\ region.
      }
      %%% In 7 steps 4850MHz 0.005 0.2 8 steps
     \label{fig05}
  \end{figure*}

The deep low resolution spectra were taken in the vicinity of the bright stars 
\object{GSC 3572--2102} and \object{GSC 3572-2103}, as well as in the south in an 
area where strong radio emission overlaps with strong \hnii\ and diffuse 
\oiii\ emission (Fig. \ref{fig01}). 
The optical spectra from areas I and II (Fig. \ref{fig04}) 
indicate emission from shock heated gas (\sii/\ha\ $>$ 0.4), 
while that from area III seems to originate from photoionized gas 
(Table \ref{sfluxes}). 
This is because we cannot exclude the possibility that the spectrum from area
III may be contaminated by emission from a nearby \HII\ region, thus 
causing the low \sii/\ha\ ratio of 0.18. 
\par
Due to the presence of strong \HII\ emission around  area I, we report the 
procedure followed with the spectral analysis. 
The first set of spectra were taken with the slit crossing the \oiii\ filament. 
However, the area along the slit and in the north of the \oiii\ filament 
contained very strong \HII\ emission. In order to take into account the diffuse
emission present in the field, the second set of spectra 
were taken with the slit shifted to the north by 6\arcmin. In this way 
the area around \dd\ $\simeq$ 46\degr00\arcmin, containing relatively weak 
\HII\ emission, was within the slit and was used as background.
It is clear that the spectral results from area I should be used under the 
assumption of constant background intensity. 
The low \oiii/\hbeta\ ratio of $\sim$6 implies a shock speed
$\sim$100 \vel\ or less, while the sulfur line ratio suggests a pre--shock cloud 
density of a few atoms per cm$^3$ (e.g. Raymond \et\ \cite{ray88}, 
Osterbrock \cite{ost89}). 
The spectrum from area II suggests electron densities below $\sim$80 \dens, 
while the strong sulfur line emission may suggest a largely neutral 
pre--shock medium. In addition, this spectrum displays very strong \oiii\ 
emission relative to \hbeta, a characteristic of shocks with incomplete 
recombination zones (Raymond \et\ \cite{ray88}). This high \oiii/\hbeta\ 
ratio of $\sim$45 further suggests shock velocities greater than 
$\sim$100 \vel. 
\par
The \ha/\hbeta\ ratio of 4.0 ($\pm$0.2) from area I results in a logarithmic 
extinction of 0.38 ($\pm$0.07) or an E(B--V) of 0.25 ($\pm$0.05), while 
in area II the logarithmic extinction of 1.1 ($\pm$0.1) 
is equivalent to an E(B--V) of 0.73 ($\pm$0.07). 
The logarithmic extinction is calculated through the relation 
c = 1/0.331 $\cdot$ log((\ha/\hbeta)$_{\rm obs}$/3) and uses the interstellar 
extinction curve of Kaler (\cite{kal76}) as implemented in the nebular 
package (Shaw and Dufour \cite{sha95}) within the IRAF software. 
The  E(B--V) is calculated with the aid of the relation 
E(B--V) = 0.664 $\cdot$ c (Kaler \cite{kal76}, Aller \cite{all84}). 
We note that the signal to noise ratios quoted in Table~\ref{sfluxes} 
do not include calibration errors which are $\sim$7\%.
\section{X--ray emission from \gsnr}
\begin {figure*}
   \resizebox{\hsize}{!}{\includegraphics{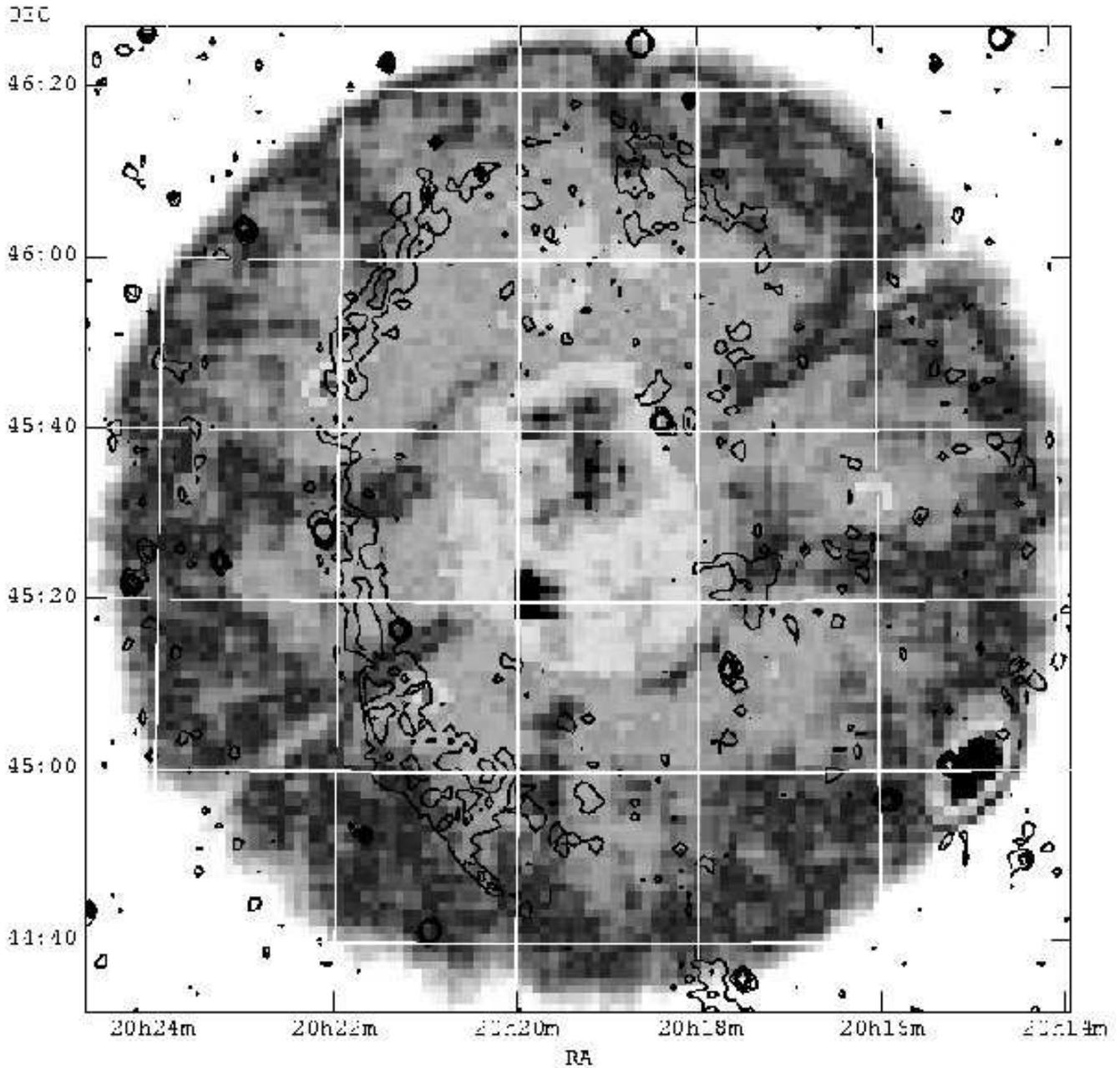}}
    \caption{ The raw soft X--ray emission detected in the 8 ks ROSAT pointed 
     observation is shown in this figure. The brightness scales linearly 
     from 2 to 24 counts/pixel and the image has been smoothed with a Gaussian 
     filter with $\sigma$=8\arcmin. 
     The radio contours at 326 MHz scale also also linearly from 0.015 to 
     0.09 Jy/beam (Rengelink \et\ \cite{ren97}).   
     } 
     \label{figref06}
  \end{figure*}

%
%The ROSAT all-sky survey data show an elliptically shaped region of X-ray emission with 
%a maximum extent of $\sim$90\arcmin, which is dominated by a relatively bright 
%bar running from the northern top to almost the southern end of the remnant. 
%The survey data also show that \gsnr\ seems to be placed in a region of low 
%X--ray surface brightness outlined by a circle of $\sim$150\arcmin\ diameter. 
%Otherwise, the Cygnus Superbubble is of similar brightness as \gsnr. 
The field of \gsnr\ was observed by ROSAT in pointed mode on 
October 26, 1993 (ROR 500286p) for 8 ks.
The image of this observation is shown in Fig. \ref{figref06}. On purpose we show just
the count/pixel image and did not correct for exposure, to illustrate where the remnant 
is highly underexposed due to the artifacts of the PSPC, clearly seen as a dark bar running from 
the lower left to the upper right of the image and other ribs, as well as a dark ring 40 arcmin wide around the 
center. Since the image is not exposure corrected the intensity close to the edge is brighter 
by a factor of about two than the brightness in the image indicates. The brightness appears quite 
clumpy and is dominated by a 
relatively bright bar off-set from the geometric centre to the north by about 9 arcmin. The bar 
measures about 24 arcmin in the north-south direction and about 18 arcmin along the east-west 
direction. 
In Fig. \ref{figref06} the PSPC image is shown along 
with contours of the radio emission at 326 MHz, which best define the 
elliptical shape of \gsnr. It is also evident from this figure that 
the X--ray emission from the remnant does not fill the PSPC 
field of view, which is supported by the all-sky image on an even wider scale. Thus it is possible 
to determine the background from the same observation. Checks with the background taken from the 
adjacent pointings have been made and the source spectrum does not change.   
\gsnr\  was also observed by the ASCA satellite on June 12, 1997. 
The ASCA X--ray telescope was pointed to the central 
area of \gsnr\ for a total of 82 ks. Given the smaller field of view of 
the GIS and SIS detectors compared to the ROSAT PSPC, this data can 
only be used for spectral analysis (see \S 4.2).
%but the bar appears to have a shape 
%slightly differing from the survey image. This is likely to be due 
%to improved counting statistics. 
%Some fraction of a surrounding shell can be seen 
%quite well in the north-east and in the south-east, although less pronounced. 
%The equivalent shell like structure seems to be missing in the west. 
%
\subsection {Image analysis}
The first and so far only results of an analysis of the ROSAT data of G 82.2+5.3 were reported by Rho \& Petre 
(\cite{rho98}), but no images were shown.  
The X--ray image looks like an SNR broken into pieces resulting into a patchy appearance. The central 
bar is the brightest component of the remnant and has led Rho \&\ Petre
(\cite{rho98}) to 
put \gsnr\  in the 
class of mixed-morphology remnants. 
First, we have searched the remnant for point sources. The ROSAT source 
detection algorithm reveals the existence of 31 point sources with a likelihood 
$>$ 10 or 4.1 $\sigma$ across the face of the remnant. Within a circle of 40\arcmin\ diameter 
centered on the apparent SNR centre 17 point--like sources are found out 
of which four sources do not have optical counterparts down to 
m$\sb{\rm B} <$ 22, with separations from known optical sources of $>$10\arcsec. 
The source existence significance of 4.1 $\sigma$ corresponds to a 
source count rate of 0.01 counts s$\sp{-1}$ which corrected for  
an absorption column density of 
N$\sb{\rm H}$ = 4.7$\times$10$\sp{21}$ cm$\sp{-2}$ 
(\S 4.2) corresponds to an unabsorbed (0.2 -- 2.4 keV) energy 
flux of $\sim$ 2$\times$10$\sp{-13}$ erg cm$\sp{-2}$ s$\sp{-1}$ for a blackbody 
source spectrum of kT$\sb{\rm{bb}}$ = 100 eV. Of course, 
with a detection significance of 4.1 $\sigma$ it can not be excluded 
that these point-like sources are actually statistical fluctuations 
above the diffuse remnant emission.
\par
\begin {figure*}
   \resizebox{\hsize}{!}{\includegraphics{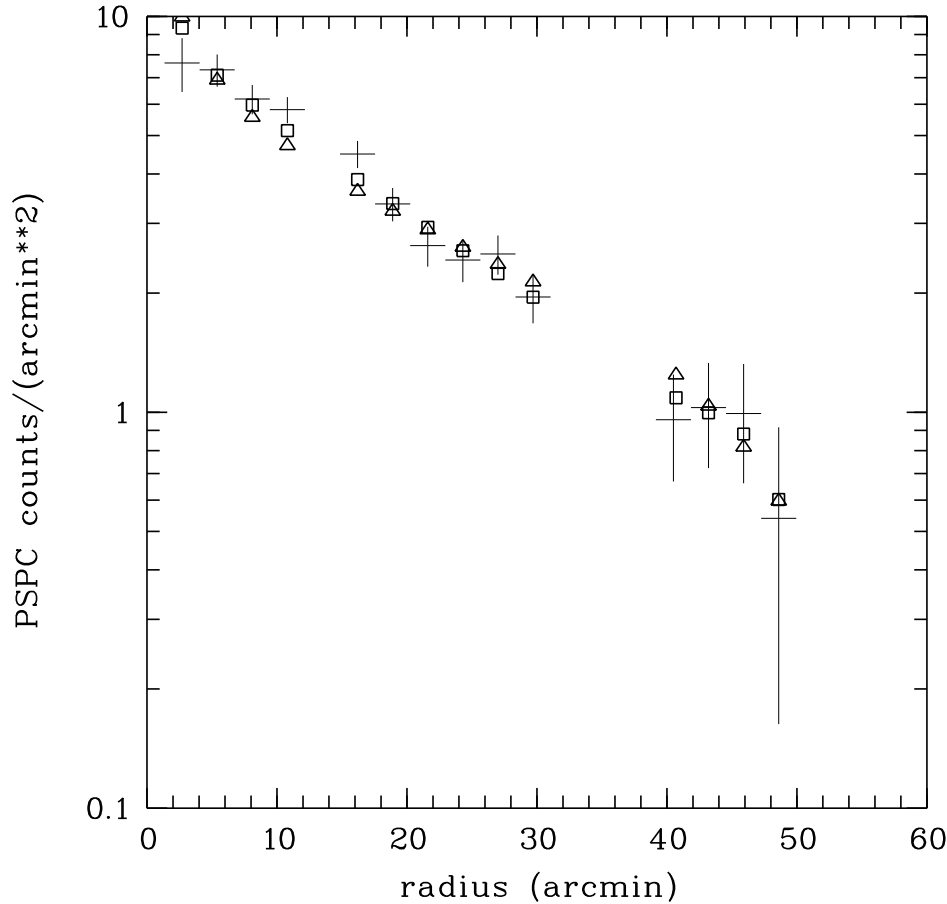}}
    \caption{The brightness profile (crosses) of G 82.2+5.3 derived from the ROSAT data is 
    shown in this figure. The open squares mark the exponential best fit 
    with some limb brightening, while the open triangles mark the power--law
    best fit. The reader is referred to \S 4.1 for more details.
     } 
     \label{density}
  \end{figure*}

Although the surface brightness is quite low, we can integrate over the azimuth 
and construct its radial profile including every PSPC count independent of 
energy. Background was subtracted, the profile was corrected for vignetting 
and the integration was done over 360$\sp{o}$ rather than just 40$\sp{o}$ which 
was used by Rho \& Petre (\cite{rho98}). 
Furthermore we centered the profile on the apparent geometric centre 
(\a\ $\simeq$ 20\h19\m07\s, \dd\ $\simeq$ 45\degr31\arcmin12\arcsec) 
of the remnant 
rather than on the brightest patch (c.f. \ref{figref06}). It turns out that the 
density profile  can be well described by an exponential law 
($\propto$ e$^{\rm -r/r_0}$) with a characteristic angular length of 
36\arcmin ($\pm$3\arcmin) up to a distance of 45\arcmin.5 ($\pm$1\arcmin.5). At
larger distances the density can be described by the same exponential law but
enhanced by a factor of 1.4$^{+0.6}_{-0.2}$, which might be interpreted as an
indication of limb brightening. The brightness profile is also consistent with a 
 density 
power law, such that the density scales inversely with the square root of the 
distance from the center. The actual data allow us to trace the density
profile out to an angular distance of $\sim$50\arcmin, which we assume to be
the radius of the remnant (Fig. \ref{density}). 
In summary, the average matter density drops by just 
about a factor of four from the center to the edge. Of course, the radial 
brightness change could also partially be produced by a radially changing 
filling factor. 
The patchiness of the surface brightness distribution is 
then a matter of local density variations and volume filling factors. 
This result is quite in contrast to the result of Rho \& Petre \cite{rho98}, who, 
however, have chosen 
a fairly small sector of the remnant (1/9 of the whole perimeter) and 
probably have centered the profile on the brightest spot. 
%
%\par
%
%{\it We mention that the surface brightness profile shows an increase of counts 
%between 42\arcmin $<r<$ 50\arcmin, which might be interpreted as an indication of 
%limb brightening. In fact, the radial surface profile can also be modeled by 
%an exponential e$^{\rm -r/r_1}$ with $r_1$ = 26\arcmin and a density increase by 
%about a factor of 2 for 42\arcmin $<r<$ 50\arcmin above the exponential.}
% 
%
\subsection{Spectral analysis}
In order to study in as much detail as possible the spectral properties
of \gsnr, we analyzed archival ASCA data and the ROSAT data. Due to the low surface brightness of the object, both the ASCA and
ROSAT PSPC data are characterized by low signal to noise, not allowing a
spatially resolved spectral analysis. Thus the analysis of the GIS data 
was restricted to the brightest region within a circle centered on 
\a\ $\simeq$ 20\h19\m14\s, \dd\ $\simeq$ 45\degr47\arcmin00\arcsec\ and
7\arcmin.4 radius. 
However, the center of the ASCA SIS detectors is offset by $\sim$6\arcmin\ 
with respect to the center of the GIS detectors. Since the SIS data 
partially overlap the GIS data, counts from a 3\arcmin$\times$3\arcmin\ square 
area were extracted from the former detectors.
\par
\begin {figure*}
   \resizebox{\hsize}{!}{\includegraphics{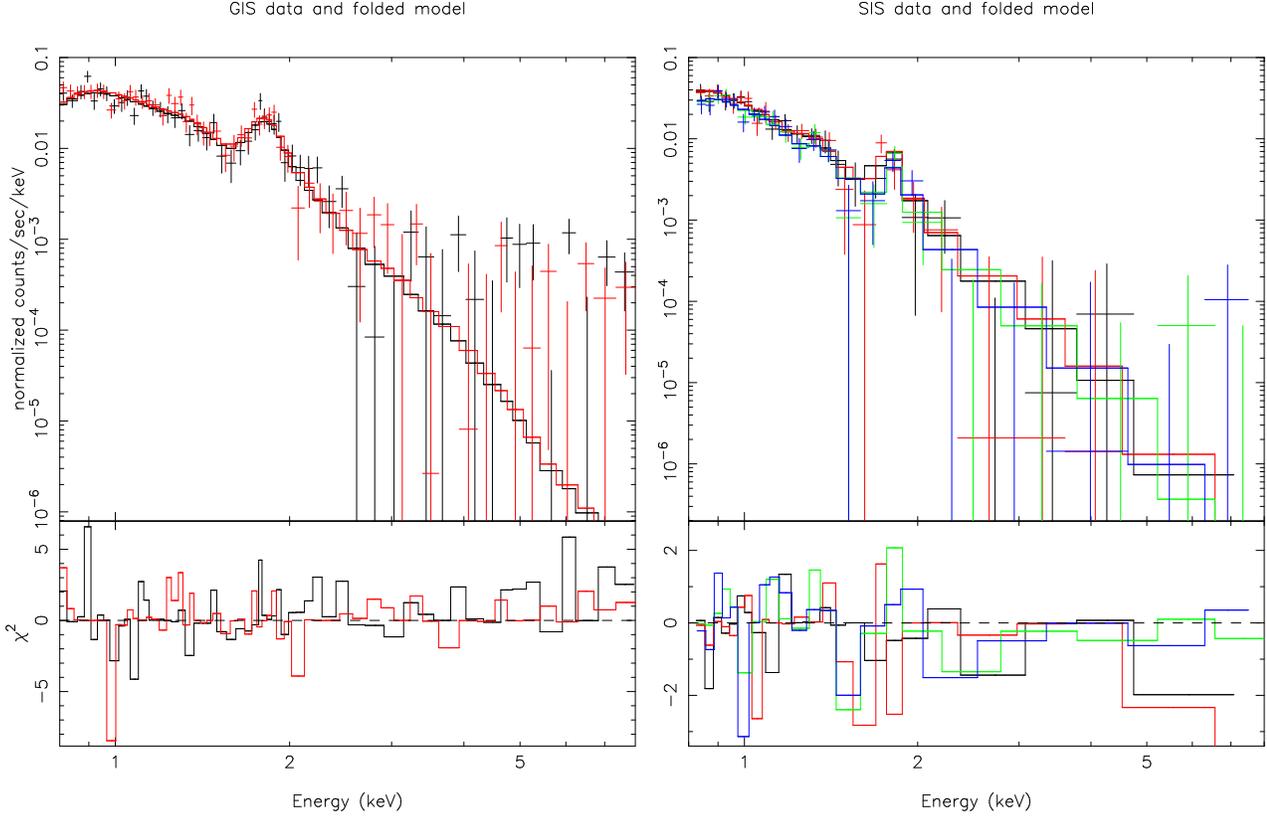}}
    \caption{The ASCA GIS2/3 data are shown in the left figure together
    with the best fit vmekal model, while the spectra from the SIS 
    detectors are shown in the right figure with the same model 
    (see \S 4.2).
     } 
     \label{gis_sis}
  \end{figure*}
\begin {figure*}
   \resizebox{\hsize}{!}{\includegraphics{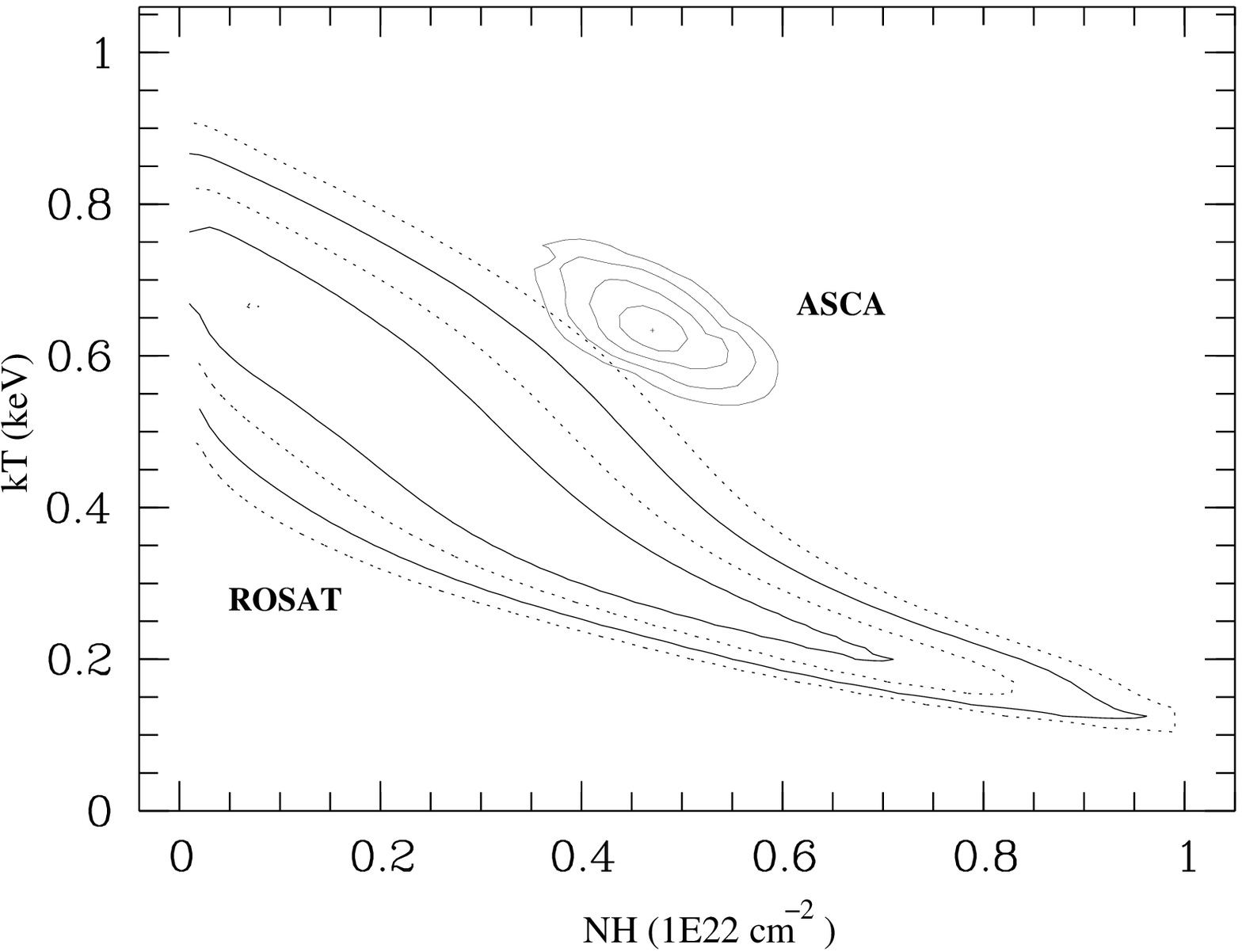}}
    \caption{Best fit and error contours (1$\sigma$ to 4$\sigma$ in steps of 
     1$\sigma$) for the temperature (kT) and the interstellar absorption column 
     density N$\sb{\rm H}$ for one-temperature vmekal models to the ROSAT PSPC 
     spectrum (lower left) and the 
     ASCA GIS/SIS spectra (upper right). Note that there is essentially no 
     overlap, indicating the need for more  advanced models. 
     } 
     \label{error_contours}
  \end{figure*}

A non--thermal component is immediately ruled out by the ASCA data since a
power--law fit is not acceptable (reduced $\chi^2$=1.74 for 197 degrees of
freedom). A mekal model fitted simultaneously to the GIS/SIS data with 
non--cosmic abundances results in a reduced $\chi^2$ of 1.15 for 196 degrees of
freedom (\dof), which is formally acceptable. 
However, the presence of a systematic trend in the residuals in the 
0.9 to 1.9 keV range suggests its rejection, while including a power--law 
component does not lead to an improved fit. 
The GIS/SIS data were also fitted with a vmekal model and a reduced 
$\chi^2$ of 0.99 for 198 \dof\ (Fig. \ref{gis_sis}) was obtained. 
The best fit parameters of this model are given in Table~\ref{bestfit}.  
Adding a second component (power--law) does not 
substantially improve the fit. The vmekal fits also constrain the abundance of 
O (1$\pm 1$), 
Ne (0.1$\sp{+0.7}\sb{-0.1}$), 
Mg (2.8$\sp{+1.3}\sb{-0.7}$), 
Si (4.0$\sp{+1.2}\sb{-1.2}$) and 
Fe (3.5$\sp{+2.2}\sb{-0.8}$); 
numbers in parentheses are 
abundances relative to solar values with 1$\sigma$ errors, which  
 shows that Mg, Si and Fe definitely differ from solar values. 
The Si line is very clearly seen in both the GIS and SIS spectra 
(c.f. fig. \ref{gis_sis}) at around 1.8 keV. This is interesting because it might mean that 
the bright patch, i.e. the near center component, is dominated by ejecta matter rather than 
circumstellar or interstellar matter. For the ROSAT analysis we used the same 
field and the same abundances with a one temperature vmekal spectrum. 
Fig. \ref{error_contours} shows the error contours relating the temperature kT and the 
interstellar absorption column density N$\sb{\rm H}$.
As the contour lines show there is basically no overlap between the ASCA and the ROSAT 
fits, except at the 4$\sigma$ level (99.99\% probability) for each of the two 
instruments. 
Clearly a one temperature model does not represent the spectrum. Obviously a component 
with a temperature too low to be observed by ASCA - we have restricted the ASCA analysis to 
energies $>$ 0.8 keV - has to be added which on the other hand is dominating the ROSAT spectrum, 
or vice versa. So we fitted the PSPC spectrum with a two component vmekal model, restricting 
the values of the high temperature,  the associated emission measure as well as 
N$\sb{\rm H}$ to the range of $\pm$1$\sigma$ around the best fit found in the ASCA data modeling. 
The best fit values for the low temperature component are given 
in Table~\ref{bestfit} as well. In summary the high temperature is found to be around 0.6 keV 
and the low temperature around 0.2 keV. The emission measure of the low temperature component 
is about three times that of 
the high temperature component. Both the N$\sb{\rm H}$ and the value of the low temperature 
are  consistent with the results of Rho \& Petre \cite{rho98} but the existence of a high temperature 
component, already indicated by a deeper analysis of the PSPC  spectrum,  
is very clearly demonstrated by adding the ASCA data to the analysis.  
\par
The emission measures of both components allow us to estimate the corresponding 
densities of the X--ray emitting gas. It is found that, for a distance of 
1.6 kpc, the average hydrogen number density of the harder component is 
$\sim$0.03 \dens, while that of the softer component is $\sim$0.05 \dens\ and both 
scale inversely with the square root of the distance to the remnant.
These numbers adopt an emission volume of $\pi\cdot$(7.4 arcmin)$\sp 2\cdot$100 arcmin, 
so that the hydrogen number density in cm$\sp{-3}$ is n$\sb{\rm H}$ = 
0.657$\times\sqrt{(\rm{EM}/\rm{d}\sb{1.6})}$, with EM the emission 
measure (c.f. Table~\ref{bestfit}).  
We note here that a change by a factor of 4 in distance would alter the
densities by only a factor of 2. It is clear that the  X--ray gas is quite
thin and that the remnant expands in an even thinner medium, if we consider 
the matter inside as being compressed by shocks. 

\par
 
As mentioned above another pointing on \gsnr\ with spectrally 
useful data has been performed. The spectral 
analysis of the PSPC pointing centered on the south--east of the remnant shows that the spectrum is 
fully consistent with that of the central part, and no significant differences can be established. 
In fact the fit with a one temperature vmekal model produces the same parameter values within 
1$\sigma$ error bars. The existence of a high temperature component cannot unambiguously 
be demonstrated, it might be 
present or might be missing. ASCA data on the field are not available.

%(Emission measure values are in units of 
%10$\sp{-14}$/(4~$\pi$d$\sp 2$)~$ {\int n_{\rm e} n_{\rm H} dV }$ throughout the paper.)% 


\section{Discussion}
The \snr\ \gsnr\ is found in a complex area in the Cygnus constellation where 
strong \HII\ emission dominates the red part of the optical spectra.
Most of the radio emission is detected in a shell which basically defines 
the boundaries of the remnant. However, a deficiency of radio emission at 
the west part of the remnant is present, at least in certain radio 
frequencies as well as in the X--rays. In this work we present the 
first flux calibrated CCD images of \gsnr\ in the emission lines of \hnii, 
\sii, \oii, and \oiii. In addition, results from a detailed imaging and 
spectral analysis of the available ROSAT and ASCA pointed observations are 
presented. 
\par
The low ionization images are dominated by the bright \HII\ region 
in the field, not allowing the identification of optical emission 
as emission from the remnant based only on morphological arguments. 
It is the ratio of the \sii\ to \hnii\ images which provides some evidence 
on the nature of the emission in this  field. 
We have also detected two filamentary structures, unknown up to now, in 
the medium ionization line of \oiii\ in the west and the east parts of the 
field observed. The two filaments are found at almost opposite sides 
and display an appreciable degree of curvature. 
The eastern filament is very well correlated with the radio emission 
at 326 and 4850 MHz (Fig. \ref{fig05}), while weak radio emission at the 
location of the western filament is only seen in the 11 cm maps 
(Velusamy \& Kundu \cite{vel74}, Wendker \cite{wen71}).
The strong differences between the low and medium ionization lines 
suggest that significant inhomogeneities and density variations in the 
pre--shock medium are present. The \ha, \nii\ and \sii\ lines being produced 
in cooler areas behind the shock front are more sensitive to 
inhomogeneities and density variations than the \oiii\ line which is produced 
closer to the shock front (e.g. Hester \cite{hes87}).  
The presence of \oiii\ emission at only certain areas of a remnant seems 
to be a common characteristic of several remnants like \object{CTB 1} 
(Fesen \et\ \cite{fes97}) and \object{G 17.4-2.3} (Boumis \et\ \cite{bou02}).
\par
The optical spectrum obtained from area I is a typical spectrum from a 
complete shock structure suggesting a column density behind the shock front of 
$\sim$10$^{18.5}$ \sdens\ (e.g. Cox \& Raymond \cite{cox85}, Raymond \et\ 
\cite{ray88}). However, the large \oiii/\hbeta\ ratio in area II can only be 
explained by an 
incomplete shock structure where the shock is traveling with a speed greater 
than 100 \vel\ in an interstellar cloud of medium density like in area I.
The swept--up column density is estimated around 10$^{18}$ \sdens\ 
(Raymond \et\ \cite{ray88}). 
The large sulfur line ratios measured in the long--slit spectra are indicative 
of low electron densities which in turn imply low pre--shock 
cloud densities (Fesen \& Kirshner \cite{fes80}). 
This is in agreement with the cloud density estimates based on the 
relative line intensities of the optical spectra and the modeling of
Raymond et al. (\cite{ray88}). 
Currently, the long--slit spectra from area III suggest emission from a
photoionized nebula. However, the complex environment of \gsnr\ does not permit 
an unambiguous determination of the nature of the optical emission in this area 
and radio observations at different frequencies are required to resolve 
this issue.
\par
In an effort to estimate the column density towards \gsnr, based on the
optical data, we use the statistical relation of 
\begin{equation}
{\rm N_{H}} = 5.4~(\pm 0.1) \cdot\ 10^{21}~{\rm E(B-V)}~{\rm cm}^{-2},
\end{equation}
given by Predehl and Schmitt (\cite{pre95}). The estimated neutral hydrogen 
column density given the observed values of E(B-V) is then in the range 
of 1--4 $\times$ 10$^{21}$ \sdens, although larger column densities cannot 
be excluded given the uncertainties in the optical extinction. 
Furthermore the locations which are accessible for N$\sb{\rm H}$ and E(B-V) 
are different. We formally know N$\sb{\rm H}$ only for the 
central part of the remnant. 
Nevertheless, the agreement with the absorbing column density derived from 
the X-ray spectrum (4.7$\times$ 10$^{21}$ \sdens) is not too bad. 
The full range of X-ray absorption of 
2.2 -- 5.0$\times$10$\sp{21}$ cm$\sp{-2}$ corresponds to  
0.4 $<$ E(B--V) $<$ 0.9. 
The E(B-V) color excess in the south is 0.94 and would formally place 
area III at a larger distance than \gsnr. 
However, the low counting statistics from area II and the complex 
environment do not allow to draw secure conclusions on the relative 
distance between the remnant and the emission in the south (area III). 
Additionally, the upper limit on the X--ray absorption does not exclude 
such a large value of E(B--V) and a correspondingly large distance.  
\par
The X--ray emission consists of a patchy disk slightly elongated in the
south--north direction, dominated by a bar-like emission close but clearly 
off-set from the geometric center. The remnant is not obviously limb brightened but a weak 
shell of emission might be present 
in the eastern hemisphere. Given the available imaging and spectral 
X--ray data, we proceed to derive some fundamental parameters of \gsnr\ 
starting with the Sedov relation. In 
this framework the remnant is considered to be in the adiabatic phase and 
estimates for the  distance d, linear diameter D, age 
$\tau$, luminosity L$\sb{\rm x}$, shock velocity v$\sb s$, pre-shock 
matter number density n$\sb 0$ and the amount of swept-up matter M$\sb{\rm{su}}$
can be derived from the angular diameter, the emission measures 
and the spectral temperatures (e.g. Pfeffermann \et\ \cite{pfe91}). 
It is interesting to note that a reasonable solution of the Sedov 
relation cannot be obtained adopting the temperature of the soft component 
(0.2 keV; see \S 4.2) as the temperature at the primary shock front.
However, the temperature of the hard component ($\sim$0.6 keV), equivalent to a
shock velocity of $\sim$720 \vel, does provide a viable solution of the Sedov
relation. Assuming a distance of 1.6 kpc, the explosion energy
would be 0.17$\cdot$10$^{51}$ erg, the age would be 13500 yr, the ISM density
would be 0.012 \dens, and the swept up mass would be 24M$_\odot$. 
If, on the other hand, we adopt the nominal value of the explosion energy 
(1$\cdot$10$^{51}$ erg), then the age would become 26700 yr, the current radius
would be 50.4 pc, the ambient density would be 0.009 \dens, the swept up mass
would be 144 M$_\odot$, and the remnant should be $\sim$3.3 kpc away from us.

\par 

The assumption of pressure equilibrium between the interstellar clouds and the 
hot gas can provide an independent estimate of the pre--shock cloud density. 
For a shock velocity of $\sim$100 \vel\ into the clouds and an ambient 
density of 0.012 \dens, a pre--shock cloud density around 1 \dens\ is estimated,
which is compatible with the estimates based on the long--slit optical spectra
(see \S 3.3). 
\par
With the above data, we can discuss the evolutionary status 
of the remnant in some more detail. Using the calculations of Cox 
\et\ (\cite{cox99}) and 
Cioffi \et\ (\cite{cio88}), it is found that the cooling times exceed  
1.3$\cdot$10$^5$ yr, i.e. a factor of 10 longer than the age estimated  
above. Furthermore, with a radius of 50 arcmin at a distance of 1.6 kpc the 
mean expansion velocity of the remnant over that very long age is just 185 \vel,
and with such a low expansion velocity one might wonder about the X-ray 
emission which has a temperature of at least 0.2 keV. Alternatively, the 
remnant might be at a greater distance, say  by a factor of two to three, which 
might fit the X-ray temperature but it is at the edge of the distance 
estimates via N$\sb{\rm H}$ and E(B-V). Furthermore the measurement of 
overabundant Si, which indicates that ejecta matter at fairly high temperatures 
still exists, does probably not favour a very old remnant. Of course it is not 
excluded that \gsnr\ has reached the radiative phase, despite the very low 
ambient density, which we stress is a direct result of the X-ray measurements 
and is independent of any assumption about the actual evolutionary status of 
the remnant. In addition, we have examined HI data 
from the Canadian Galactic Plane survey (CGPS -- Taylor \et\ \cite{tay03}). 
Unfortunately, due to its high galactic latitude of $\sim$5\degr\ only 
emission from the eastern areas of the remnant was recorded in this survey. 
So, at least for this part of the remnant, we were not able to identify 
any features that would point to an HI shell. 

\par
The low ambient density may result from the interaction of the stellar wind 
of the progenitor star with its environment since a typical wind blown bubble 
is characterized by a radius of $\sim$30 pc, a density of $\sim$ 0.01 \dens\ 
and a hot interior (e.g. Castor \et\ \cite{cas75}, Weaver \et\ \cite{wea77}). 
The low surface X--ray brightness of the annulus around \gsnr, if associated 
with it, would support the suggestion of the wind blown bubble in which 
the progenitor star of \gsnr\ exploded and its remnant is expanding.   
\par
The ASCA spectra of \gsnr\ show that the central X--ray emission is of thermal
origin, while the ROSAT images show a azimuthally averaged brightness profile which is 
consistent with a slow exponential decrease of the matter density by a factor of about four from 
the center to the edge. 
In principle, the evaporating cloud model proposed by White and Long
(\cite{whi91}) or the model of Maciejewski \et\  (\cite{maj99}), where a remnant 
evolves in a stratified medium, can be of relevance to W 63. This model could 
probably explain the patchiness and the cloudy structure.  
Furthermore, heat conduction 
may apply in order to smooth the temperature distribution in the 
interior of the remnant and make it look isothermal 
(e.g. Chevalier \cite{che99}). However, with the currently available data we 
cannot claim the existence of such mechanisms. 
\par
At last, there is an interesting aspect which comes to mind after the analysis 
of the ROSAT PSPC spectra. There may be the possibility that the X-ray 
emitting gas we see is just a minor fraction of the total matter inside the 
remnant. If there were gas at temperatures below about 0.05 keV, it would have 
been missed by the X-ray instruments, even if the emission measure would be 
higher by two to three orders in magnitude, basically because the intersteller 
absorption column density is so high, and this holds for all the mixed 
morphology remnants listed (N$\sb{\rm H}>3.8\times$10$\sp{21}$ cm$\sp{-2}$) by 
Rho \& Petre (\cite{rho98}).
\section{Conclusions}
We have obtained deep flux calibrated images of \gsnr. The low ionization images
are dominated by the \HII\ region in the field but the medium ionization line
of \oiii\ provides the first clear picture of the remnant. 
It displays an elliptical shape with filamentary structures in the east 
and west areas. The correlation between the optical and radio emission 
supports their association. Long--slit spectra
suggest both complete and incomplete recombination zones, shock velocities  
around 100 \vel, and low electron densities.
The X-ray emission region appears to  largely be embedded in the optical and radio emission. 
The X--ray surface brightness is quite patchy,
missing obvious limb brightening and is dominated by a bright
bar--like emission region which is off-set from the geometric center by
$\sim$9\arcmin. The X-ray emission is definitely thermal and requires two 
temperatures of 0.2 keV and 0.63 keV. The bright bar region shows  
overabundant Mg, Si and Fe, which might indicate still 
radiating ejecta matter. 
The azimuthally averaged radial surface profile is consistent with the matter 
density changing with distance r from the center $\propto$e$^{\rm -r/r_0}$ with 
a characteristic angular length of 36\arcmin; but it is also consistent with 
an r$\sp{-1/2}$ density profile. The matter inside the remnant is
quite likely structured like a porous cloudy medium. The average matter
density derived from the X-ray surface brightness is 
$\sim$0.05$\times{\rm d\sb{1.6}\sp{-0.5}}$ 
and $\sim$0.03$\times{\rm d\sb{1.6}\sp{-0.5}}$ for the two spectral components,
respectively. It could be slightly higher with filling factors significantly 
less than 1. 
From the X-ray and optical measurements we have determined some parameters 
of the SNR, but the evolution of the remnant into the mixed--morphology 
state is still difficult to understand. If we apply the 'Sedov analysis' 
approach, which would assume the remnant to be in the adiabatic phase, 
the remnant is at a distance between 1.6 kpc and 3.3 kpc with an age 
of 13.5 and 26.7 kyrs. These results appear reasonable, particularly 
because the distance is consistent with that determined by Rosado and 
Gonzales (\cite{ros81}). But the remnant shows neither the expected density 
profile nor a strong limb brightening for a constant ambient density on which
the 'Sedov analysis' is based. In addition,  
we determine the temperature only at the center and not at the limb 
which is what is required for the application of the 'Sedov analysis'. 
On the other hand, the assumption that the remnant is in the radiative phase 
entering the shell--forming phase is in conflict with the very low density of 
the X-ray emitting plasma and the excessive long radiative cooling time 
involved. Although the low matter density itself does not completely rule out 
that the remnant has reached the radiative phase it would imply an even 
lower density, greater age and much larger distance at the edge 
of the upper limits obtained from N$\sb{\rm H}$ and E(B-V). These conclusions 
are primarily based on the low density of the X-ray plasma but the optical 
filaments demonstrate the existence of matter of pre-shock densities of one 
or a few atoms \dens\ and shock velocities of ~100 \vel. Whether these 
filaments are located around the perimeter of the remnant or inside of it is
not clear. But matter at these densities and velocities of $>$100 \vel\ 
should 
emit soft X-rays or EUV radiation. The fact that we do not see such 
radiation puts an upper limit on the temperature and the emission measure, or 
electron density, of this component in order to stay 'undetected' in the 
PSPC observations which depend on N$\sb{\rm{H}}$. Adopting the observed value 
of N$\sb{\rm{H}}$ the pre-shock electron density could be as high as
1 \dens\ for log(T) = 5.8 or kT = 55  eV. Interestingly, the radiative 
cooling time is ~25 kyrs, and an HI shell might not have formed yet. 
We have not found any evidence for an HI shell, but this is not conclusive 
as the full area around the remnant has not been imaged so far. 
Even higher electron densities up to 20 \dens\  with 
log(T) = 5.5 are possible for the soft X-rays/EUV  to escape detection but the 
radiative cooling time gets as short as 10 kyrs, and the probability to have 
caught a remnant in such a brief transition phase gets increasingly lower. 
If most of the dense matter were predominantly located around the perimeter 
of the remnant it might represent the shock heated matter of parts of the 
inner edge of a stellar wind--blown bubble, which might still be larger than 
the remnant, indicated by the low-level X-ray emission surrounding 
the remnant as seen in the ROSAT All-sky survey. If the supernova had 
a stellar wind--blown bubble, the low electron density derived from the X-ray 
measurements is easy to understand as it matches the prediction of densities 
in such bubbles. The temperature and the X-ray surface brightness distribution 
observed in the remnant are then a consequence of the temperature and 
matter density distributions in the bubble prior to the supernova shock, both 
of which change substantially with distance from the progenitor star 
(Weaver et al. \cite{wea77}).
\begin{acknowledgements}
\end{acknowledgements}
We thank the referee for careful and productive comments and pointing us 
to the ASCA data; we are grateful to Emi Miyata for her advice on analysing 
the ASCA data. 
Skinakas Observatory is a collaborative project of the University of
Crete, the Foundation for Research and Technology-Hellas and
the Max-Planck-Institut f\"ur Extraterrestrische Physik.
This research has made use of data obtained through the High Energy 
Astrophysics Science Archive Research Center Online Service, 
provided by the NASA/Goddard Space Flight Center. 
The data from the 
Canadian Galactic Plane Survey were obtained from the Canadian 
Astronomy Data Centre (where F.M. is a guest user) which is operated 
by the Herzberg Institute of Astrophysics of the National 
Research Council Canada.

%

%----------------
\vfill\eject
%
%--------------------------------------------------------
%  \begin {figure*}
 %  \resizebox{\hsize}{!}{\includegraphics{fig3727.ps}}
%    \caption{ The \HII\ emission is somewhat suppressed in the low ionization 
%     lines of \oii\ but still the shape of \gsnr\ is not clearly outlined.  
%     The image has been smoothed to suppress the residuals from the imperfect 
%     continuum subtraction. Shadings run linearly from 
%     0 to 100 $\times$ \flux. 
%     } 
%     \label{fig03}
%  \end{figure*}
%
\end{document}